\documentclass[aps,onecolumn,a4paper,superscriptaddress,longbibliography,nofootinbib,notitlepage]{revtex4-1}
\usepackage[cmex10]{amsmath}
\usepackage{graphicx,epic,eepic,epsfig,latexsym,amssymb,verbatim,color,revsymb}

\newcommand{\abs}[1]{\left\lvert#1\right\rvert} % absolute value: single vertical bars
\usepackage[a4paper,total={6.5in,9in}]{geometry}
\usepackage[english]{babel}
\usepackage[T1]{fontenc}
\usepackage{mathrsfs}
\usepackage{enumerate}
\usepackage{graphicx}
\usepackage{mathtools}
\usepackage[export]{adjustbox}
\usepackage{microtype}
\usepackage{MnSymbol}
\usepackage{lipsum}

\usepackage[colorlinks = true,
            linkcolor = red,
            urlcolor  = blue,
            citecolor = blue,
            anchorcolor = red]{hyperref}

\newtheorem{theorem}{Theorem}

\newtheorem{lemma}[theorem]{Lemma}
 
\newtheorem{definition}[theorem]{Definition}  
\newtheorem{remark}[theorem]{Remark}  

\newcommand\qedsymbol{$\blacksquare$}

\newlength{\blank}
\newenvironment{proof-of}[1][{\hspace{-\blank}}]{{\medskip\noindent\textit{Proof~{#1}.\ }}}{\hfill\qedsymbol}
\newenvironment{proof}{{\medskip\noindent\textit{Proof.\ }}}{\hfill\qedsymbol}

\newcommand{\Tr}{{\operatorname{Tr}}}

\newcommand{\id}{{\operatorname{id}}}

\newcommand{\1}{\openone}
\newcommand{\ket}[1]{|#1\rangle}

\newcommand{\proj}[1]{|#1\rangle\!\langle #1|}

\newcommand{\cC}{{\mathcal{C}}}
\newcommand{\cD}{{\mathcal{D}}}
\newcommand{\cE}{{\mathcal{E}}}
\newcommand{\cR}{{\mathcal{R}}}
\newcommand{\cT}{{\mathcal{T}}}

\newcommand{\KI}{{\text{KI}}}

\begin{document}

\title{General Mixed State Quantum Data Compression \protect\\
       with and without Entanglement Assistance} 

\author{Zahra Baghali Khanian}
\affiliation{F\'{\i}sica Te\`{o}rica: Informaci\'{o} i Fen\`{o}mens Qu\`{a}ntics, %
Departament de F\'{\i}sica, Universitat Aut\`{o}noma de Barcelona, 08193 Bellaterra (Barcelona), Spain}
\affiliation{ICFO---Institut de Ci\`{e}ncies Fot\`{o}niques, \protect\\%
Barcelona Institute of Science and Technology, 08860 Castelldefels, Spain}
\email{zbkhanian@gmail.com}

\author{Andreas Winter}
\affiliation{F\'{\i}sica Te\`{o}rica: Informaci\'{o} i Fen\`{o}mens Qu\`{a}ntics, %
Departament de F\'{\i}sica, Universitat Aut\`{o}noma de Barcelona, 08193 Bellaterra (Barcelona), Spain}
\affiliation{ICREA---Instituci\'o Catalana de Recerca i Estudis Avan\c{c}ats, %
Pg.~Lluis Companys, 23, 08010 Barcelona, Spain} 
\email{andreas.winter@uab.cat}

\date{16 December 2019}

%\author{Zahra Baghali Khanian and Andreas Winter}
%\title{Compression of Quantum Source Ensembles \protect\\ with Entanglement Assistance} 

%\author{%
%  \IEEEauthorblockN{Zahra Baghali Khanian\IEEEauthorrefmark{1}\IEEEauthorrefmark{2}}
%  \IEEEauthorblockA{\IEEEauthorrefmark{1}%
%                    ICFO\\
%                    Barcelona Institute of Technology\\
%                    08860 Castelldefels, Spain\\
%                    Email: zbkhanian@gmail.com}
%  \and
%  \IEEEauthorblockN{ }
%  \IEEEauthorblockA{\IEEEauthorrefmark{2}%
%                    Grup d'Informaci\'{o} Qu\`{a}ntica\\
%                    Departament de F\'{\i}sica\\
%                    Universitat Aut\`{o}noma de Barcelona\\
%                    08193 Bellaterra (Barcelona), Spain}
%  \and
%  \IEEEauthorblockN{Andreas Winter\IEEEauthorrefmark{2}\IEEEauthorrefmark{3}}
%  \IEEEauthorblockA{\IEEEauthorrefmark{3}%
%                    ICREA\\
%                    Pg.~Lluis Companys, 23\\
%                    08010 Barcelona, Spain\\
%                    Email: andreas.winter@uab.cat}
%}

\begin{abstract}
We consider the most general (finite-dimensional) quantum 
mechanical information source, which is given by a quantum system 
$A$ that is correlated with a reference system $R$. The task is to 
compress $A$ in such a way as to reproduce the joint source state 
$\rho^{AR}$ at the decoder with asymptotically high fidelity. This 
includes Schumacher's original quantum source coding problem of a 
pure state ensemble and that of a single pure entangled state, as 
well as general mixed state ensembles. 
Here, we determine the optimal compression rate (in qubits per 
source system) in terms of the Koashi-Imoto decomposition of the 
source into a classical, a quantum, and a redundant part. The same 
decomposition yields the optimal rate in the presence of unlimited 
entanglement between compressor and decoder, and indeed the full 
region of feasible qubit-ebit rate pairs. 
%
%
%
%In this work, we consider the most general finite dimensional source in the realm of quantum mechanics, that is a quantum system $A$ with Hilbert space $\mathcal{H}_A$ of finite dimension which is correlated with a reference system $R$ with Hilbert space $\mathcal{H}_R$ of finite dimension with the overall state $\rho^{AR}$ where the reference does not necessarily purify the source. In this work, we find the optimal rate for the compression of this source which is the von Neumann entropy of systems $C Q$ in the Koashi-Imoto (KI) decomposition of the state $\rho^{AR}$. Moreover, we find the optimal rate when the compression task is assisted by free entanglement shared between the encoder and decoder. We show that the optimal compression rate is reduced in this case by half of the entropy of the classical system $C$ in the KI decomposition of the state.
\end{abstract}

\maketitle

\section{What is a quantum source?}
A quantum source is a quantum system together correlations with a \emph{reference system}.
A criterion of how well a source is reproduced in a communication 
task is to measure how well the correlations are preserved with 
the reference system. Without correlation, the information does not make sense 
because a known quantum state without correlations can be reproduced at the 
destination without any communication.

%
%\textcolor{green}{(AW: In particular the explanation of why 
%you need an explicit ("physical") reference system to even say  
%what information is. I don't know if there is a concrete reference, 
%but it definitely was something around, and very much in Ben 
%Schumacher's work.)} 
%
%Therefore, this definition of a quantum source involve a measure to gauge  the correlations and ..
%Without correlation the information does not make sense because a known quantum system 
%without correleations can be make in the destination without any communication.  
%

To elaborate more on these notions, consider the source that Schumacher defined in his 1995 paper \cite{Schumacher1995,Jozsa1994_1} as an ensemble of pure states $\{p(x),\ket{\psi_x}^{A} \}$,
where the source generates the state $\ket{\psi_x}$ with probability $p(x)$. The figure of merit for the encoding-decoding process is to keep the decoded quantum states \emph{on average} very close to the original states with respect to the fidelity, where the average is taken over the probability distribution $p(x)$. 
By basic algebra one can show that this is equivalent to preserving the classical-quantum state 
$\rho^{AX}=\sum_x p(x) \proj{\psi_x}^A \otimes \proj{x}^X$, where system $A$ is the quantum system to be compressed. 
Another source model that Schumacher considered was the purification of the source ensemble, 
that is the state $\ket{\psi}^{AR}=\sum_x \sqrt{p(x)}\ket{\psi_x}^{A} \ket{x}^R$, where the 
figure of merit for the encoding-decoding process was to preserve the pure state correlations with the 
reference system $R$ by maintaining a high fidelity between the decoded state and $\psi$. 
He showed that both definitions lead to the same compression rate, namely, 
the von Neumann entropy of the source $S(A)_{\rho} = S(\rho^A)$, where 
$\rho^A = \Tr_R \rho^{AR}$. Incidentally, the full proof of optimality in the first 
model, without any additional restrictions on the encoder, had to wait until \cite{Barnum1996}
(see also \cite{Horodecki1998});
the strong converse, i.e. the optimality of the entropy rate even for constant error 
bounded away from $1$, was eventually given in \cite{Winter:PhD}.

Another example of a quantum source is the mixed state source considered by Horodecki \cite{Horodecki1998} and Barnum \emph{et al.} \cite{Barnum2001}, and finally solved by Koashi and Imoto \cite{KI2001},
where the source is defined as an ensemble of mixed states $\{p(x),\rho_x^{A} \}$. Preserving 
these mixed quantum states, on average, in the process of encoding-decoding, the task is 
equivalent to preserving the state $\rho^{AX}=\sum_x p(x) \rho_x^A \otimes \proj{x}^X$, that is 
the quantum system $A$ together with its correlation with the classical reference system $X$.    

The reference system is not usually considered in the description of classical 
information theory tasks, but arguably it is conceptually necessary in quantum 
information. This is because it allows us to present the figure of merit quantifying 
the decoding error as operationally accessible, for example via the probability of
passing a test in the form of a measurement on the combined $AR$-system. This point 
is made eloquently in the early work of Schumacher on quantum information transmission
\cite{Schumi1996,Barnum1998}. 

In this work, we consider the most general finite-dimensional source in the realm of 
quantum mechanics, namely a quantum system $A$ that is correlated with a reference system 
$R$ in an arbitrary way, described by the overall state $\rho^{AR}$. In particular, the 
reference does not necessarily purify the source, nor is it assumed to be classical. 
%
%The ensemble source and the pure source considered by Schumacher as well as the mixed ensemble source considered by Koashi and Imoto in \cite{KI2001} are special cases of this model where the references are either classical systems in the ensemble models or a purifying system in the pure source of Schumacher. 
%
The ensemble source and the pure source defined by Schumacher are special cases of this model, 
where the reference is a classical system in the former and a purifying system in the latter. 
So is the source considered by Koashi and Imoto in \cite{KI2001}, where the reference system 
is classical, too. 

Understanding the compression of the source $\rho^{AR}$ has paramount importance in the 
field of quantum information theory and unifies all the models that have been considered 
in the literature. Schumacher's pure source model in a sense is the most stringent model 
because it requires preserving the correlations with a purifying reference system which 
implies that the correlations with any other reference system is preserved which follows 
from the fact that the fidelity is non-decreasing under quantum channels. 
However, the converse is not necessarily true: if in a compression task the parties are 
required to preserve the correlations with a given reference system which does not purify 
the source state, they might be able to compress more efficiently compared to the scenario 
where the reference system purifies the source. This is exactly what we show in this paper: 
we characterise the gap precisely depending on the reference system.

We find the optimal trade-off between the quantum and entanglement rates of the compression 
which are in terms of a decomposition of the state $\rho^{AR}$ introduced in \cite{Hayden2004}. 
This decomposition is a generalization of the decomposition introduced by Koashi and Imoto
for a set of quantum states in \cite{KI2002}, so when the reference system is classical, 
the quantum rate reduces to the rate derived by Koashi and Imoto.  
We show the optimality of the rates with a new converse proof which is based on the decoupling of the environment systems of the encoding and decoding operations from the decoded systems and gives us an insight into how general mixed states are processed in an encoding-decoding task. 
%Whereas the converse for the special case of the classical reference system in \cite{KI2001} 
%is based on replacing the source with an equivalent but pure state ensemble source. 
Our results also cover the entanglement assisted compression task considered in \cite{ZBK2019} when the side information system is trivial, as well as the entanglement assisted version of the Koashi-Imoto compression.

\medskip

The structure of the paper is as follows. In the reminder of this section, we introduce the 
notation that we use throughout the paper. 
In Sec.~\ref{sec:The Compression task}, we
rigorously define the task of the asymptotic compression of the source $\rho^{AR}$, where as for the communication purposes, we let the encoder and decoder share initial entanglement, 
and the encoder sends the compressed information to the decoder through a noiseless quantum channels. 
In Sec.~\ref{sec:The optimal rate region}, we first introduce the Koashi-Imoto decomposition 
of the state $\rho^{AR}$, and then in Theorem~\ref{theorem:complete rate region} we state the main result of this paper, that is the optimal rate region for the compression of the source
in terms of the trade-off between the entanglement and quantum rates, then we prove the achievability of the rates in the same section, but we leave the converse proofs for the subsequent sections which need more involved machinery. In Sec.~\ref{sec:converse}, we define two functions which emerge in the converse proofs, and in Lemma~\ref{lemma:J_epsilon Z_epsilon properties} we state some important properties of these functions which then we use to prove the tight asymptotic converse bounds of Theorem~\ref{theorem:complete rate region}. 
We prove Lemma~\ref{lemma:J_epsilon Z_epsilon properties} in Sec.~\ref{sec: Proof of Lemma}. 
Finally, in Sec.~\ref{sec:Discussion} we discuss our results and some related open problems.

\medskip
\textbf{Notation.} 
Quantum systems are associated with (finite dimensional) Hilbert spaces $A$, $R$, etc.,
whose dimensions are denoted by $|A|$, $|R|$, respectively. Since it is clear from the context, we slightly abuse the notation and let $Q$ denote both a quantum system and a quantum rate. 
%We identify states with their density operators, and we use the notation $\phi= \ketbra{\phi}{\phi}$ as the density operator of the pure state vector $\ket{\phi}$. 
The von Neumann entropy is defined as $S(\rho) = - \Tr\rho\log\rho$ 
(throughout this paper, $\log$ denotes by default the binary logarithm).
%and its inverse function $\exp$, unless otherwise stated, is also to basis $2$).
The conditional entropy and the conditional mutual information, $S(A|B)_{\rho}$ and $I(A:B|C)_{\rho}$,
respectively, are defined in the same way as their classical counterparts: 
\begin{align*}
  S(A|B)_{\rho}   &= S(AB)_\rho-S(B)_{\rho}, \text{ and} \\ 
  I(A:B|C)_{\rho} &= S(A|C)_\rho-S(A|BC)_{\rho} \\
                    &= S(AC)_\rho+S(BC)_\rho-S(ABC)_\rho-S(C)_\rho.
\end{align*}
The fidelity between two states $\rho$ and $\xi$ is defined as 
\(
 F(\rho, \xi) = \|\sqrt{\rho}\sqrt{\xi}\|_1 
                = \Tr \sqrt{\rho^{\frac{1}{2}} \xi \rho^{\frac{1}{2}}},
\) 
with the trace norm $\|X\|_1 = \Tr|X| = \Tr\sqrt{X^\dagger X}$.
It relates to the trace distance in the following well-known way \cite{Fuchs1999}:
\begin{equation}
  1-F(\rho,\xi) \leq \frac12\|\rho-\xi\|_1 \leq \sqrt{1-F(\rho,\xi)^2}.
\end{equation}

\section{The compression task}
\label{sec:The Compression task}
We will consider the information theoretic limit of
many copies of the source $\rho^{AR}$, i.e.~$\rho^{A^n R^n} = \left(\rho^{AR}\right)^{\otimes n}$.
We assume that the encoder, Alice, and the decoder, Bob, have initially a maximally 
entangled state $\Phi_K^{A_0B_0}$ on registers $A_0$ and $B_0$ (both of dimension $K$).
The encoder, Alice, performs the encoding compression operation 
$\mathcal{C}:A^n A_0 \longrightarrow M $ on the system $A^n$ and her part $A_0$ of the entanglement, which is a quantum channel,
i.e.~a completely positive and trace preserving (CPTP) map. 
Notice that as functions CPTP maps act on the operators (density matrices) over 
the respective input and output Hilbert spaces, but as there is no risk of confusion,
we will simply write the Hilbert spaces when denoting a CPTP map.
Alice's encoding operation produces the state $\sigma^{M B_0 R^n}$ with $M$ and $B_0$ as the compressed system of Alice and Bob's part of the entanglement, respectively.
The dimension of the compressed system is without loss of 
 generality not larger than  the dimension of the
original source, i.e. $|M| \leq  \abs{A}^n$. 
We call $\frac1n \log K$ and $\frac1n \log|M|$ the \emph{entanglement
rate} and \emph{quantum
rate} of the compression protocol, respectively.
The system $M$ is then sent to Bob via a noiseless quantum channel, who performs
a decoding operation $\mathcal{D}:M B_0 \longrightarrow \hat{A}^n$ on the system 
$M$ and his part of the entanglement $B_0$.
We say the encoding-decoding scheme has \emph{fidelity} $1-\epsilon$, or \emph{error} $\epsilon$, if 
\begin{align}
  \label{eq:fidelity criterion}
  F\left( \rho^{A^n R^n },\xi^{\hat{A}^n R^n} \right)  
          \geq 1-\epsilon,  
\end{align}
where $\xi^{\hat{A}^n R^n}=\left((\mathcal{D}\circ\mathcal{C})\otimes \id_{R^n}\right) \rho^{A^n R^n }$.
Moreover, we say that $(E,Q)$ is an (asymptotically) achievable rate pair if for all $n$
there exist codes such that the fidelity converges to $1$, and
the entanglement and quantum rates converge to $E$ and $Q$, respectively.
The rate region is the set of all achievable rate pairs, as a subset of 
$\mathbb{R}_{\geq 0}\times\mathbb{R}_{\geq 0}$. 

According to Stinespring's theorem \cite{Stinespring1955}, a CPTP map 
$\cT: A \longrightarrow \hat{A}$ can be dilated to an isometry $U: A \hookrightarrow \hat{A} E$ 
with $E$ as an environment system, called an isometric extension of a CPTP map, such that 
$\cT(\rho^A)=\Tr_E (U \rho^A U^{\dagger})$. 
Therefore, the encoding and decoding operations are can in general be viewed as 
isometries $U_{\cE} : A^n A_0 \hookrightarrow M W$ and
$U_{\cD} : M B_0 \hookrightarrow \hat{A}^n V$, respectively, 
with the systems $W$ and $V$ as the environment systems
of Alice and Bob, respectively. 

We say a source $\omega^{BR}$ is equivalent to a source $\rho^{AR}$ if there are
CPTP maps $\cT:A \longrightarrow B$ and $\cR:B \longrightarrow A$ in both directions
taking one to the other: 
\begin{align} \label{def: equivalent sources}
    \omega^{BR}=(\cT \otimes \id_R) \rho^{AR} \text{ and } 
    \rho^{AR}=(\cR \otimes \id_R) \omega^{BR}.
\end{align}
The rate regions of equivalent sources are the same, because any achievable rate pair for 
one source is achievable for the other source as well. This follows from the fact that for
any code $(\cC,\cD)$ of block length $n$ and error $\epsilon$ for $\rho^{AR}$, 
concatenating the encoding and decoding operations with $\cT$ and $\cR$, i.e. letting
$\cC'=\cC\circ\cR^{\otimes n}$ and $\cD'=\cT^{\otimes n}\circ\cD$, we get a code 
of the same error $\epsilon$ for $\omega^{BR}$. Analogously we can turn a code for 
$\omega^{BR}$ into one for $\rho^{AR}$.

\section{The qubit-ebit rate region}
\label{sec:The optimal rate region}
The idea behind the compression of the source $\rho^{AR}$ is based on a decomposition 
of this state introduced in \cite{Hayden2004}, which is a generalization of the decomposition
introduced by Koashi and Imoto in \cite{KI2002}. Namely, for any set of quantum states $\{ \rho_x\}$, 
there is a unique decomposition of the Hilbert space describing
the structure of CPTP maps which preserve the set $\{ \rho_x^A\}$. This idea was generalized 
in \cite{Hayden2004} for a general mixed state $\rho^{AR}$ describing the structure of 
CPTP maps acting on system $A$ which preserve the overall state $\rho^{AR}$. 
This was achieved by showing that any such map preserves the set of all possible
states on system $A$ which can be obtained by measuring system $R$, and 
conversely any map preserving the set of all possible
states on system $A$ obtained by measuring system $R$, preserves the state $\rho^{AR}$,
thus reducing the general case to the case of classical-quantum states 
\[
  \rho^{AY} = \sum_y q(y) \rho_y^A \otimes \proj{y}^Y
            = \sum_y \Tr_R \rho^{AR}(\1_A\otimes M_y^R) \otimes \proj{y}^Y, 
\]
which is the ensemble case considered by Koashi and Imoto. As a matter of fact, 
looking at the algorithm presented in \cite{KI2002} to compute the decomposition,
it is enough to consider an informationally complete POVM $(M_y)$ on $R$, with 
no more than $|R|^2$ many outcomes.
The properties of this decomposition are stated in the following theorem.

\begin{theorem}[\cite{KI2002,Hayden2004}]
\label{thm: KI decomposition}
Associated to the state $\rho^{AR}$, there are Hilbert spaces $C$, $N$ and $Q$
and an isometry $U_{\KI}:A \hookrightarrow C N Q$ such that:
\begin{enumerate}
  \item The state $\rho^{AR}$ is transformed by $U_{\KI}$ as
    \begin{equation}
      \label{eq:KI state}
      (U_{\KI}\otimes \1_R)\rho^{AR} (U_{\KI}^{\dagger}\otimes \1_R)
        = \sum_j p_j \proj{j}^{C} \otimes \omega_j^{N} \otimes \rho_j^{Q R}
        =:\omega^{C N Q R},
    \end{equation}
    where the set of vectors $\{ \ket{j}^{C}\}$ form an orthonormal basis for Hilbert space 
    $C$, and $p_j$ is a probability distribution over $j$. The states $\omega_j^{N}$ and 
    $\rho_j^{Q R}$ act  on the Hilbert spaces $N$ and $Q \otimes R$, respectively.

  \item For any CPTP map $\Lambda$ acting on system $A$ which leaves the state $\rho^{AR}$ 
    invariant, that is $(\Lambda \otimes \id_R )\rho^{AR}=\rho^{AR}$, every associated 
    isometric extension $U: A\hookrightarrow A E$ of $\Lambda$ with the environment system 
    $E$ is of the following form
    \begin{equation}
      U = (U_{\KI}\otimes \1_E)^{\dagger}
            \left( \sum_j \proj{j}^{C} \otimes U_j^{N} \otimes \1_j^{Q} \right) U_{\KI},
    \end{equation}
    where the isometries $U_j:N \hookrightarrow N E$ satisfy 
    $\Tr_E [U_j \omega_j U_j^{\dagger}]=\omega_j$ for all $j$.
    The isometry $U_{KI}$ is unique (up to trivial change of basis of the Hilbert spaces 
    $C$, $N$ and $Q$). Henceforth, we call the isometry $U_{\KI}$ and the state 
    $\omega^{C N Q R}=\sum_j p_j \proj{j}^{C} \otimes \omega_j^{N} \otimes \rho_j^{Q R}$ 
    the Koashi-Imoto (KI) isometry and KI-decomposition of the state $\rho^{AR}$, respectively. 

  \item In the particular case of a tripartite system $CNQ$ and a state $\omega^{CNQR}$ already 
    in Koashi-Imoto form (\ref{eq:KI state}), property 2 says the following:
    For any CPTP map $\Lambda$ acting on systems $CNQ$ with 
    $(\Lambda \otimes \id_R )\omega^{CNQR}=\omega^{CNQR}$, every associated 
    isometric extension $U: CNQ\hookrightarrow CNQ E$ of $\Lambda$ with the environment system 
    $E$ is of the form
    \begin{equation}
      U = \sum_j \proj{j}^{C} \otimes U_j^{N} \otimes \1_j^{Q},
    \end{equation}
    where the isometries $U_j:N \hookrightarrow N E$ satisfy 
    $\Tr_E [U_j \omega_j U_j^{\dagger}]=\omega_j$ for all $j$.
\end{enumerate} 
\end{theorem}

According to the discussion at the end of Sec. \ref{sec:The Compression task}, the 
sources $\rho^{AR}$ and $\omega^{C N Q R}$ are equivalent because there are the isometry 
$U_{\KI}$ and the reversal CPTP map $\cR: C N Q \longrightarrow A$, which reverses the 
action of the KI isometry, such that:
\begin{align}
    \omega^{C N Q R}&= (U_{\KI}\otimes \1_R)\rho^{AR} (U_{\KI}^{\dagger}\otimes \1_R) \nonumber \\
    \rho^{AR}&=(\cR \otimes \id_R)\omega^{C N Q R}=(U_{\KI}^{\dagger }\otimes \1_R) \omega^{C N Q R} (U_{\KI}\otimes \1_R)+\Tr [(\1_{C N Q }-\Pi_{C N Q})\omega^{C N Q}]\sigma,
\end{align}
where $\Pi_{C N Q}=U_{\KI}U_{\KI}^{\dagger}$ is the projection onto the subspace 
$U_{\KI}A \subset C \otimes N \otimes Q$, and $\sigma$ is an arbitrary state acting on $A\otimes R$.
Henceforth we assume that the source is $\omega^{C N Q R}$, which is convenient because
our main result is expressed in terms of the systems $C$ and $Q$. Notice that
the source $\omega^{C N Q R}$ is in turn equivalent to $\omega^{C Q R}$,
a fact we will exploit in the proof.

%For the source $\omega^{C N Q R}$, the system $N$ is decoupled from the reference system $R$ given 
%the classical information $j$ in the classical system $C$.
%Thus, $N$ is a redundant system in the sense that its state
%is a function of $j$, so by having access to $j$ this system can be generated locally without a need to access the reference system $R$. This is the core idea behind the compression of this source.
%
%In essence, we show that it is necessary and sufficient to compress the systems $C Q$,
%and system $N$ is a redundant system which can be generated locally at the encoder and decoder site.
Moreover, since the information in $C$ is classical, we can reduce the 
compression rate even more if the sender and receiver share  
entanglement, by using dense coding of $j$. In the following
theorem we show the optimal qubit-ebit rate tradeoff for the compression of the source $\rho^{AR}$. 

%\medskip
\begin{theorem}
  \label{theorem:complete rate region}
  For the compression of the source $\rho^{AR}$, all asymptotically achievable entanglement and
  quantum rate pairs $(E,Q)$ satisfy
  \begin{align*}
    Q   &\geq S(CQ)_{\omega}-\frac{1}{2}S(C)_{\omega},\\
    Q+E &\geq  S(CQ)_{\omega}, 
  \end{align*}
  where the entropies are with respect the KI decomposition of the state $\rho^{AR}$, i.e. 
 the state $\omega^{C N Q R}$.
  Conversely, all the rate pairs satisfying the above inequalities are asymptotically achievable.
\end{theorem}

\begin{remark}
This theorem implies that the optimal asymptotic quantum rates for the compression of 
the source $\rho^{AR}$ with and without entanglement assistance are 
$S(CQ)_{\omega}-\frac{1}{2}S(C)_{\omega}$ and $S(CQ)_{\omega}$ qubits, respectively,
and $\frac{1}{2}S(C)_{\omega}$ ebits of entanglement 
are sufficient and necessary in the entanglement assisted case. 
\end{remark}

\begin{remark}
If in the compression task the parties were required to preserve the correlations with a purifying reference system, then due to Schumacher compression the optimal qubit rate would be $S(A)_{\rho}=S(CNQ)_{\omega}$. However, Theorem~\ref{theorem:complete rate region} shows that 
the parties can compress more if they are only required to preserve the correlations with a mixed state reference. This gap can be strictly positive if the redundant system $N$ is mixed given the classical information $j$ in system $C$, that is $S(CNQ)_{\omega}-S(CQ)_{\omega}=S(N|CQ)_{\omega}>0$. 
\end{remark}

\begin{figure}[ht] 
  \includegraphics[width=0.8\textwidth]{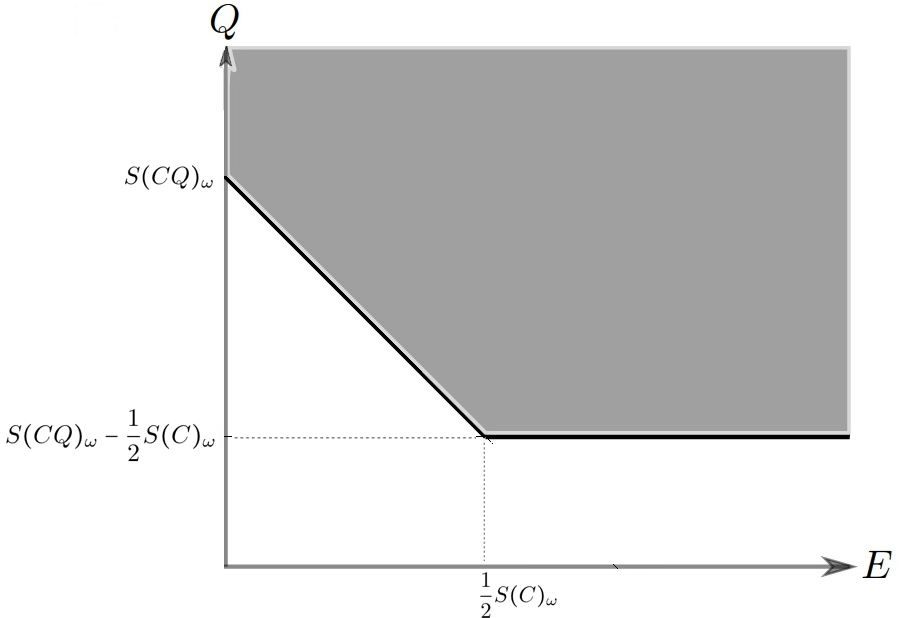} 
  \caption{The achievable rate region of the entanglement and quantum rates.}
  \label{fig:rate region}
\end{figure}

\begin{proof}
We start with the achievability of these rates. The converse proofs need more tools, so we will 
leave them to the subsequent sections. 
Looking at Fig. \ref{fig:rate region}, it will be enough to 
prove the achievability of the corresponding corner points $(E,Q)=(0,S(CQ)_{\omega})$ 
and $(E,Q)=(\frac{1}{2}S(C)_{\omega},S(CQ)_{\omega}-\frac{1}{2}S(C)_{\omega})$ 
for the unassisted and entanglement assisted cases, respectively. This is because
by definition (and the time-sharing principle) the rate region is convex and 
upper-right closed.
Indeed, all the points on the line $Q+E = S(CQ)_{\omega}$ for 
$Q \geq S(CQ)_{\omega}-\frac{1}{2}S(C)_{\omega}$ are achievable because one 
ebit can be distributed by sending a qubit. 
All other rate pairs are achievable by resource wasting. 
The rate region is depicted in Fig.~\ref{fig:rate region}.

As we discussed, we can assume that the source is 
$(\omega^{C N Q R})^{\otimes n}=\omega^{C^n N^n Q^n R^n}$. To achieve the point $(0,S(CQ)_{\omega})$, 
Alice traces out the redundant part $N^n$ of the source, to get the state $\omega^{C^n Q^n R^n}$ and applies Schumacher compression to send the systems $C^n Q^n$ to Bob. Since the 
Schumacher compression preserves the purification of the systems $C^n Q^n$, it preserves the state $\omega^{C^n Q^n R^n}$ as well. To be more specific, let $\Lambda_S$ denote the composition of the encoding and decoding operations for the Schumacher compression of the state
$\ket{\omega}^{C^n Q^n R^n {R'}^n}$ where the system ${R'}^n$ is a purifying reference system which of course the parties do not have access to. The Schumacher compression preserves the following fidelity on the left member of the equation, therefore it preserves the fidelity on the right member:
\begin{align}%\label{eq:Schumacher fidelity}
   1-\epsilon 
     \leq F \left({\omega}^{C^n Q^n R^n {R'}^n} ,(\Lambda_S \otimes \id_{R^n {R'}^n}) {\omega}^{C^n Q^n R^n {R'}^n}\right) 
     \leq F \left({\omega}^{C^n Q^n R^n } ,(\Lambda_S \otimes \id_{R^n } ){\omega}^{C^n Q^n R^n }\right), \nonumber
\end{align}
where the inequality is due to monotonicity of the fidelity under partial trace. 
The rate achieved by this scheme is $S(CQ)_{\omega}$. After applying this scheme, 
Bob has access to the systems $\hat{C}^n \hat{Q}^n$, which is correlated with the reference 
system $R^n$:
\begin{align*}
  \zeta^{\hat{C}^n \hat{Q}^n R^n}=(\Lambda_S \otimes \id_{R^n } ){\omega}^{C^n Q^n R^n }.
\end{align*}
Then, to reconstruct the system $N^n$, Bob applies the CPTP map 
$\mathcal{N}:C Q  \longrightarrow C  N Q$ to each copy, which acts as follows:
\begin{align}%\label{CPTP map: cT}
  \mathcal{N}(\rho^{CQ})=\sum_j (\proj{j}^{C} \otimes \1_{Q}) \rho^{CQ} (\proj{j}^{C} \otimes\1_{Q}) \otimes \omega_j^{N}.\nonumber   
\end{align}
%
%Then, he applies the reversal CPTP map $\cR: C N Q \longrightarrow A$ %with isometric extension $U_{\cR}: C N Q \hookrightarrow A Y$ and  $Y$ is the environment system of this reverse map.
%to reconstruct the original state:
%\begin{align*}
%        \xi^{\hat{A}^n R^n}=(\cR^{\otimes n} \otimes \id_{R^n })(\cT^{\otimes n} \otimes \id_{R^n } )\zeta^{\hat{C}^n \hat{Q}^n R^n},
%\end{align*}
%where the reversal map $\cR$ reverses the action of the KI isometry $U_{\KI}$ and acts as the following:  
%\begin{align}\label{CPTP map: cR}
%    \cR(\rho)=U_{\KI}^{\dagger } \rho U_{\KI}+\Tr [(\1_{C N Q }-\Pi_{C N Q})\rho]\sigma,
%\end{align}
%where $\Pi_{C N Q}=U_{\KI}U_{\KI}^{\dagger }$ is a projection onto the Hilbert space $C \otimes N \otimes Q$, and $\sigma$ is a state acting on the Hilbert space $A$.
%
This map satisfies the fidelity criterion of Eq.~(\ref{eq:fidelity omega})  because of monotonicity of the fidelity under CPTP maps:
\begin{align}\label{eq:fidelity omega}
    1-\epsilon &\leq F \left({\omega}^{C^n Q^n R^n } ,\zeta^{\hat{C}^n \hat{Q}^n R^n}\right) \nonumber\\
    & \leq F \left((\mathcal{N}^{\otimes n} \otimes \id_{R^n } ){\omega}^{C^n Q^n R^n } ,(\mathcal{N}^{\otimes n} \otimes \id_{R^n } ) \zeta^{\hat{C}^n \hat{Q}^n R^n}\right) \nonumber \\
    &= F \left({\omega}^{C^n N^n Q^n R^n } ,\tau^{\hat{C}^n \hat{N}^n \hat{Q}^n R^n}\right).
\end{align}

To achieve the point $(\frac{1}{2}S(C)_{\omega},S(CQ)_{\omega}-\frac{1}{2}S(C)_{\omega})$, 
Alice applies dense coding to send the classical system $C^n$ to Bob which requires 
$\frac{n}{2}S(C)_{\omega}$ ebits of initial entanglement and $\frac{n}{2}S(C)_{\omega}$ 
qubits \cite{Bennett1992}. When both Alice and Bob have access to system $C^n$, 
Alice can send the quantum system $Q^n$ to Bob by applying Schumacher compression, which 
requires sending $nS(Q|C)$ qubits to Bob.
Therefore, the overall qubit rate is 
$\frac{1}{2}S(C)_{\omega}+S(Q|C)=S(CQ)_{\omega}-\frac{1}{2}S(C)_{\omega}$.
\end{proof}

\section{Converse}
\label{sec:converse}
In this section, we will provide the converse bounds for the qubit rate $Q$ and the sum 
rate $Q+E$ of Theorem~\ref{theorem:complete rate region}.
We obtain these bounds based on the structure of the CPTP maps which preserve the source state
$\omega^{CNQR}$. Namely, according to Theorem~\ref{thm: KI decomposition} the CPTP maps 
acting on systems $CNQ$, which preserve the state $\omega^{CNQR}$, act only on the 
redundant system $N$. This implies that the environment systems of such CPTP maps are 
decoupled from systems $Q R$ given the classical information $j$ in the classical system $C$.
This gives us an insight into the structure of the  encoding-decoding maps, which preserve 
the overall state \emph{asymptotically} intact. 

To proceed with the proof, we first define two functions that emerge in the converse bounds.
Then, we state some important properties of these functions in 
Lemma~\ref{lemma:J_epsilon Z_epsilon properties} which we will use to
compute the tight asymptotic converse bounds.
%The functions $Z_\epsilon(\omega)$ and $J_\epsilon(\omega)$ emerge in the converse proof of the entanglement assisted model. In the un-assisted model, only $J_\epsilon(\omega)$ emerges.  

\begin{definition}
  \label{def:J_epsilon Z_epsilon}
  For the KI decomposition  
  $\omega^{C N Q R}=\sum_{j} p_j \proj{j}^{C}\otimes \omega_j^{N} \otimes \rho_{j}^{Q R}$
  of the state $\rho^{AR}$ and $\epsilon \geq 0$, define
  \begin{align*}
    J_\epsilon(\omega) &:=  
        \max I(\hat{N} E:\hat{C}\hat{Q}|C')_\tau 
                  \text{ s.t. } U:C N Q \rightarrow \hat{C} \hat{N} \hat{Q} E
                  \text{ is an isometry with } 
                  F( \omega^{C N Q R},\tau^{\hat{C} \hat{N} \hat{Q}R})  \geq 1- \epsilon,  \\
    Z_\epsilon(\omega) &:=  
        \max S(\hat{N} E|C')_\tau 
          \text{ s.t. } U:C N Q \rightarrow \hat{C} \hat{N} \hat{Q} E 
          \text{ is an isometry with } 
          F( \omega^{C N Q R},\tau^{\hat{C} \hat{N} \hat{Q}R})  \geq 1- \epsilon,  
\end{align*}
where
\begin{align*}
  \omega^{C N Q  R C'} 
     &=\sum_{j} p_j \proj{j}^{C}\otimes \omega_j^{N} \otimes \rho_{j}^{Q R} \otimes \proj{j}^{C'}, \\
  \tau^{\hat{C} \hat{N} \hat{Q} ER C'}
     &= (U \otimes \1_{RC'}) \omega^{C N Q R C'}    (U^{\dagger} \otimes \1_{RC'}), \\
  \tau^{\hat{C} \hat{N} \hat{Q} R}
     &=\Tr_{E C'} [ \tau^{\hat{C} \hat{N} \hat{Q}  ER C'}].
\end{align*}
\end{definition}
In this definition, the dimension of the environment is w.l.o.g. bounded as $|E| \leq (|C||N||Q|)^2$
because the input and output dimensions of the channel are fixed as $|C||N||Q|$; 
hence, the optimisation is of a continuous function over a compact domain, so we have a 
maximum rather than a supremum.

\begin{lemma}
  \label{lemma:J_epsilon Z_epsilon properties}
  The functions $Z_\epsilon(\omega)$ and $J_\epsilon(\omega)$ have the following properties:
  \begin{enumerate}
    \item They are non-decreasing functions of $\epsilon$. 
    \item They are concave in $\epsilon$.
    \item They are continuous for $\epsilon \geq 0$. 
    \item For any two states $\omega_1^{C_1 N_1 Q_1 R_1}$ and $\omega_2^{C_2 N_2 Q_2 R_2}$ and for $\epsilon \geq 0$,
    \begin{align*}
        &J_{\epsilon}(\omega_1 \otimes \omega_2) \leq  J_{\epsilon}(\omega_1) +J_{\epsilon}(\omega_2),\\  
        &Z_{\epsilon}(\omega_1 \otimes \omega_2) \leq  Z_{\epsilon}(\omega_1) +Z_{\epsilon}(\omega_2).  
    \end{align*}
    \item At $\epsilon=0$, $Z_0(\omega) =S(N|C)_\omega$ and $J_0(\omega) =0$.
  \end{enumerate}
\end{lemma}

The proof of this lemma follows in the next section. Now we show how it is
used to prove the converse (optimality) of Theorem \ref{theorem:complete rate region}.
As a guide to reading the subsequent proof, we remark that in Eqs.~(\ref{eq:converse_Q_1}) 
and (\ref{eq:converse_Q+E_2}), the environment systems $VW$ of the encoding-decoding 
operations appear in the terms $I(\hat{N}^n VW : \hat{C}^n  \hat{Q}^n| {C'}^n)$ and 
$S(\hat{N}^n VW | {C'}^n)$, which are bounded by  
the functions $J_{\epsilon}(\omega^{\otimes n})$ and $Z_{\epsilon}(\omega^{\otimes n})$, 
respectively.
As stated in point 4 of Lemma~\ref{lemma:J_epsilon Z_epsilon properties}, these functions 
are sub-additive, so basically we can single-letterize the terms appearing in the converse.
Moreover, from point 3 of Lemma~\ref{lemma:J_epsilon Z_epsilon properties}, we know that
these functions are continuous for $\epsilon \geq 0$; therefore, the limit points of these 
functions are equal to the values of these functions at $\epsilon=0$.
When the fidelity is equal to 1 ($\epsilon=0$), the structure of the CPTP maps 
preserving the state $\omega^{C N Q R}$ in Theorem~\ref{thm: KI decomposition} 
implies that $J_{0}(\omega)=0$ and  $Z_{0}(\omega)=S(N|C)_{\omega}$, as stated 
in point 5 of Lemma~\ref{lemma:J_epsilon Z_epsilon properties}. Thereby, we conclude 
the converse bounds in Eqs.~(\ref{eq:converse_Q_asymptotics}) and (\ref{eq:converse_Q+E_asymptotics}).

\begin{proof-of}[of Theorem \ref{theorem:complete rate region} (converse)]
We first get the following chain of inequalities considering the process of the 
decoding of the information: 
\begin{align}
  nQ+S(B_0)&\geq S(M)+S(B_0)      \label{eq:converse_decoding_1_1}  \\
           &\geq S(M B_0)         \label{eq:converse_decoding_1_2} \\
            %&= S(\hat{A}^n V)     \label{eq:converse_decoding_1_3} \\
           &= S(\hat{C}^n \hat{N}^n \hat{Q}^n V)     \label{eq:converse_decoding_1_4} \\
           &= S(\hat{C}^n\hat{Q}^n) + S(\hat{N}^n V| \hat{C}^n  \hat{Q}^n)     \label{eq:converse_decoding_1_5} \\
           &\geq nS(C  Q) + S(\hat{N}^n V| \hat{C}^n  \hat{Q}^n)    -n \delta(n,\epsilon)    \label{eq:converse_decoding_1_6}\\
           &\geq nS(C  Q) + S(\hat{N}^n V| \hat{C}^n  \hat{Q}^n {C'}^n)    -n \delta(n,\epsilon)    \label{eq:converse_decoding_1_7}\\
           &= nS(C  Q) + S(\hat{N}^n V| \hat{C}^n  \hat{Q}^n {C'}^n) - S(\hat{N}^n V| {C'}^n)+ S(\hat{N}^n V| {C'}^n)-n \delta(n,\epsilon)  \nonumber\\
           &=  nS(C  Q) - I(\hat{N}^n V : \hat{C}^n  \hat{Q}^n| {C'}^n)+ S(\hat{N}^n V| {C'}^n)  -n \delta(n,\epsilon)\nonumber\\
            %&\geq  nS(C  Q) - I(\hat{N}^n V : \hat{C}^n  \hat{Q}^n| {C'}^n)+ S(\hat{N}^n V| {C'}^n)-n \delta(n,\epsilon)\nonumber\\
           &\geq nS(C Q) - I(\hat{N}^n VW : \hat{C}^n  \hat{Q}^n| {C'}^n)+ S(\hat{N}^n V| {C'}^n)-n \delta(n,\epsilon) \label{eq:converse_decoding_1}
\end{align}
where Eq.~(\ref{eq:converse_decoding_1_1}) follows because the entropy of a system 
is bounded by the logarithm of the dimension of that system;
Eq.~(\ref{eq:converse_decoding_1_2}) is due to sub-additivity of the entropy;
Eq.~(\ref{eq:converse_decoding_1_4}) follows because the decoding isometry 
$U_{\cD}:M B_0 \hookrightarrow \hat{C}^n \hat{N}^n \hat{Q}^n V$ does not change the entropy;
Eq.~(\ref{eq:converse_decoding_1_5}) is due to the chain rule;
Eq.~(\ref{eq:converse_decoding_1_6}) follows from the decodability: the 
output state on systems $\hat{C}^n \hat{Q}^n$ is $2\sqrt{2\epsilon}$-close 
to the original state $C^n  Q^n$ in trace norm; then the inequality follows 
by applying the Fannes-Audenaert inequality 
\cite{Fannes1973,Audenaert2007}, where 
$\delta(n,\epsilon)=\sqrt{2\epsilon} \log(|C||Q|) + \frac1n h(\sqrt{2\epsilon})$;
Eq.~(\ref{eq:converse_decoding_1_7}) is due to strong sub-additivity of the entropy,
and system $C'$  is a copy of classical system $C$;
Eq.~(\ref{eq:converse_decoding_1}) 
follows from data processing inequality where $W$ is the environment system 
of the encoding isometry $U_{\cE}:C^n N^n Q^n A_0 \hookrightarrow M W$.

Moreover, considering the process of encoding the information, $Q$ is bounded as follows:
\begin{align}
  nQ &\geq S(M)  \nonumber \\                   
     &\geq S(M|W  {C'}^n)           \label{eq:converse_encoding_1_1} \\
     &=    S(M W  {C'}^n) -S(W  {C'}^n) \label{eq:converse_encoding_1_2} \\
     %&=    S(A^n  A_0 {C'}^n)-S(W  {C'}^n) \label{eq:converse_encoding_1_3} \\
     %&=    S(A^n  {C'}^n)+S(A_0)-S(W {C'}^n) \label{eq:converse_encoding_1_4} \\
     %&=    S(A^n   {C'}^n)+S(A_0)-S({C'}^n)-S(W |{C'}^n) \label{eq:converse_encoding_1_5} \\
     &=    S(C^n N^n Q^n A_0  {C'}^n)-S(W {C'}^n) \label{eq:converse_encoding_1_4} \\
     &=    S(C^n N^n Q^n {C'}^n)+S(A_0)-S(W {C'}^n) \label{eq:converse_encoding_1_5} \\
     &=    S(C^n N^n Q^n  {C'}^n)+S(A_0)-S({C'}^n)-S(W  |{C'}^n) \label{eq:converse_encoding_1_6} \\
     &=    S(C^n N^n Q^n)+S(A_0)-S({C'}^n)-S(W |{C'}^n) \label{eq:converse_encoding_1_7} \\
     &=    nS(C Q)+nS(N|C Q)+S(A_0)-nS(C')-S(W  |{C'}^n) \label{eq:converse_encoding_1_8} \\
     &=    nS(C   Q )+nS(N |C)+S(A_0)-nS(C')-S(W |{C'}^n),  \label{eq:converse_encoding_1}
\end{align}
where Eq.~(\ref{eq:converse_encoding_1_1}) is due to sub-additivity of the entropy;
%as explained in the previous paragraph, $Y$ is the environment of the isometry $U_{\cR}: C N Q \hookrightarrow A Y$ which reverses the action of the KI isometry $U_{\KI}$.
%
Eq.~(\ref{eq:converse_encoding_1_2}) is due to the chain rule;
Eq.~(\ref{eq:converse_encoding_1_4}) follows because the encoding isometry 
$U_{\cE}:C^n N^n Q^n A_0 \hookrightarrow M W$ does not the change the entropy; 
Eq.~(\ref{eq:converse_encoding_1_5}) follows because the initial entanglement $A_0$ 
is independent from the source;
Eq.~(\ref{eq:converse_encoding_1_6}) is due to the chain rule;
%
%Eq.~(\ref{eq:converse_encoding_1_6}) follows because the reversal isometry $U_{\cR}: C N Q \hookrightarrow A Y$ does not change the entropy.
%
Eq.~(\ref{eq:converse_encoding_1_7}) follows because $C'$ is a copy of the system $C$, 
so $S(C'|C  N Q)=0$;
Eq.~(\ref{eq:converse_encoding_1_8}) is due to the chain rule and the fact that the entropy is additive
for product states;
Eq.~(\ref{eq:converse_encoding_1}) follows because conditional on system $C$ 
the system $N$ is independent from system $Q$. 

Now, we add Eqs.~(\ref{eq:converse_decoding_1}) and 
(\ref{eq:converse_encoding_1}); the entanglement terms $S(A_0)$ and $S(B_0)$ cancel out,
and by dividing by $2n$ we obtain
\begin{align}
  Q \!&\geq S(C Q)-\frac{1}{2}S(C)\! +\frac{1}{2}S(N |C)\!-\frac{1}{2n} I(\hat{N}^n VW  : \hat{C}^n  \hat{Q}^n| {C'}^n)+ \frac{1}{2n}S(\hat{N}^n V| {C'}^n)\!-\frac{1}{2n}S(W |{C'}^n)\!-\frac{1}{2} \delta(n,\epsilon)  \nonumber\\
  &\geq S(C Q)-\frac{1}{2}S(C) +\frac{1}{2}S(N |C)-\frac{1}{2n} I(\hat{N}^n VW : \hat{C}^n  \hat{Q}^n| {C'}^n)- \frac{1}{2n}S(\hat{N}^n VW | {C'}^n) -\frac{1}{2} \delta(n,\epsilon)  \label{eq:converse_Q_1}\\
  &\geq S(C Q)-\frac{1}{2}S(C) +\frac{1}{2}S(N |C)-\frac{1}{2n}J_{\epsilon}(\omega^{\otimes n})-\frac{1}{2n} Z_{\epsilon}(\omega^{\otimes n})- \frac{1}{2} \delta(n,\epsilon) \label{eq:converse_Q_2} \\ 
  &\geq S(C Q)-\frac{1}{2}S(C) +\frac{1}{2}S(N |C)-\frac{1}{2}J_{\epsilon}(\omega)-\frac{1}{2} Z_{\epsilon}(\omega)- \frac{1}{2} \delta(n,\epsilon), \label{eq:converse_Q}   
\end{align} 
%where in the first line $S(C')$ is replaced by $S(C)$ because system $C'$ is a copy of classical system $C$, therefore they both have the same entropy. 
where Eq.~(\ref{eq:converse_Q_1}) follows from strong sub-additivity of the entropy, 
$S(\hat{N}^n V| {C'}^n)+S(\hat{N}^n V| W  {C'}^n)\geq 0$; 
Eq.~(\ref{eq:converse_Q_2}) follows from Definition~\ref{def:J_epsilon Z_epsilon};
Eq.~(\ref{eq:converse_Q}) is due to point 4 of Lemma~\ref{lemma:J_epsilon Z_epsilon properties}.

In the limit of $\epsilon \to 0$ and 
$n \to \infty $, the qubit rate is thus bounded by
\begin{align}\label{eq:converse_Q_asymptotics}
  Q &\geq S(C Q)-\frac{1}{2}S(C) +\frac{1}{2}S(N |C)-\frac{1}{2}J_{0}(\omega)-\frac{1}{2} Z_{0}(\omega) \nonumber\\
    &= S(C Q)-\frac{1}{2}S(C),
\end{align}
where the equality follows from point 5 of Lemma~\ref{lemma:J_epsilon Z_epsilon properties}. 

Moreover, from Eq.~(\ref{eq:converse_decoding_1}) we have:
\begin{align}
   nQ+S(B_0)&= nQ+nE     \nonumber  \\
            &\geq nS(C Q) - I(\hat{N}^n VW : \hat{C}^n  \hat{Q}^n| {C'}^n)+ S(\hat{N}^n V| {C'}^n)-n \delta(n,\epsilon) \nonumber \\
            &\geq nS(C Q) - I(\hat{N}^n VW : \hat{C}^n  \hat{Q}^n| {C'}^n)-n \delta(n,\epsilon) \label{eq:converse_Q+E_2} \\
            &\geq nS(C Q) - J_{\epsilon}(\omega^{\otimes n})-n \delta(n,\epsilon) \label{eq:converse_Q+E_3} \\
            &\geq nS(C Q) - nJ_{\epsilon}(\omega)-n \delta(n,\epsilon), \label{eq:converse_Q+E}  
\end{align}
where Eq.~(\ref{eq:converse_Q+E_2}) follows because the entropy conditional on a 
classical system is positive, $S(\hat{N}^n V| {C'}^n) \geq 0$;
Eq.~(\ref{eq:converse_Q+E_3}) follows from Definition~\ref{def:J_epsilon Z_epsilon};
Eq.~(\ref{eq:converse_Q+E})  is due to point 4 of Lemma~\ref{lemma:J_epsilon Z_epsilon properties}. 

In the limit of $\epsilon \to 0$ and 
$n \to \infty $, we thus obtain the following bound on the rate sum:
\begin{align}
  Q+E \geq S(CQ) - J_{0}(\omega)
      =    S(CQ) \label{eq:converse_Q+E_asymptotics},  
\end{align}
where the equality follows from point 5 of Lemma~\ref{lemma:J_epsilon Z_epsilon properties}. 
\end{proof-of}

\begin{remark}
Our lower bound on $Q+E$ in Eq. (\ref{eq:converse_Q+E_asymptotics}) reproduces the 
result of Koashi and Imoto \cite{KI2001} for the case of a classical-quantum source
$\rho^{AX} = \sum_x p(x) \rho_x^A \otimes \proj{x}^X$. This is because a code with
qubit-ebit rate pair $(Q,E)$ gives rise to a compression code in the sense of 
Koashi and Imoto using a rate of qubits $Q+E$ and no prior entanglement, simply by
first distributing $E$ ebits and then using the entanglement assisted code. 

It is worth noting that conversely, Eq. (\ref{eq:converse_Q+E_asymptotics}) can 
be obtained from the Koashi-Imoto result, as follows. Any good code for $\rho^{AR}$
is automatically a good code for the classical-quantum source of mixed states
\[
  \rho^{AY} = \sum_y q(y) \rho_y^A \otimes \proj{y}^Y
            = \sum_y \Tr_R \rho^{AR}(\1_A\otimes M_y^R) \otimes \proj{y}^Y, 
\]
for any POVM $(M_y)$ on $R$, simply by the monotonicity of the fidelity 
under CPTP maps. As discussed before, by choosing an informationally complete
measurement, the KI-decomposition of the ensemble $\{q(y),\rho_y^A\}$ is 
identical to that of $\rho^{AR}$ in Theorem \ref{thm: KI decomposition}.
Thus the unassisted qubit compression rate of $\rho^{AY}$ and of $\rho^{AR}$
are lower bounded by the same quantity, the right hand side of Eq. (\ref{eq:converse_Q+E_asymptotics}).
\end{remark}

\section{Proof of Lemma~\ref{lemma:J_epsilon Z_epsilon properties}}
\label{sec: Proof of Lemma}

\begin{enumerate}
%%%%%%%%%%%%%%%%%% 1 %%%%%%%%%%%%%%%%%%%%
\item The definitions of the functions $J_{\epsilon}(\omega)$ and $Z_{\epsilon}(\omega)$ 
  directly imply that they are non-decreasing functions of $\epsilon$.

%%%%%%%%%%%%%%%%%% 2 %%%%%%%%%%%%%%%%%%%%
\item We first prove the concavity of $Z_{\epsilon}(\omega)$. 
  Let $U_1:C N Q \hookrightarrow \hat{C} \hat{N} \hat{Q} E$ and 
  $U_2:C N Q \hookrightarrow \hat{C} \hat{N} \hat{Q} E$ be the isometries attaining the 
  maximum for $\epsilon_1$ and $\epsilon_2$, respectively, which act as 
  follows on the purification $\ket{\omega}^{C N Q R C' R'}$ of the previously
  introduced state $\omega^{C N Q R C'}$:
  \[
    \ket{\tau_1}^{\hat{C} \hat{N} \hat{Q} E R C' R'}
        =(U_1 \otimes \1_{R C' R'}) \ket{\omega}^{C N Q R C' R'}
    \text{ and }
    \ket{\tau_2}^{\hat{C} \hat{N} \hat{Q} E R C' R'}
        =(U_2 \otimes \1_{R C' R'}) \ket{\omega}^{C N Q R C' R'},   
  \]
  where $\Tr_{R'}[\proj{\omega}^{C N Q R C' R'}]=\omega^{C N Q R C'}$. 
  For $0\leq \lambda \leq 1$, define the isometry 
  $U_0:C N Q \hookrightarrow \hat{C} \hat{N} \hat{Q} E F F'$ which acts as 
  \begin{equation}
    \label{eq: isometry U in convexity}
    U_0 := \sqrt{\lambda} U_1 \otimes \ket{11}^{FF'} + \sqrt{1-\lambda} U_2 \otimes \ket{22}^{FF'},
  \end{equation}
  where systems $F$ and $F'$ are qubits, and
  which leads to the state
  \begin{equation}
    (U_0 \otimes \1_{R C' R'}) \ket{\omega}^{C N Q R C' R'}
      = \sqrt{\lambda}\ket{\tau_1}^{\hat{C} \hat{N} \hat{Q} E R C' R'} \ket{11}^{FF'}
        + \sqrt{1-\lambda}\ket{\tau_2}^{\hat{C} \hat{N} \hat{Q} E R C' R'} \ket{22}^{FF'}.   
  \end{equation}
  Then, $U_0$ defines its state $\tau$. for which the reduced state on the systems 
  $\hat{C} \hat{N} \hat{Q}  R C'$ is 
  \begin{align} \label{eq: tau in convexity proof}
    \tau^{\hat{C} \hat{N} \hat{Q} R C'} 
      =\lambda \tau_1^{\hat{C} \hat{N} \hat{Q} R C'}+ (1-\lambda) \tau_2^{\hat{C} \hat{N} \hat{Q} R C'}. 
  \end{align}  
  Therefore, the fidelity for the state $\tau$ is bounded as follows:
  \begin{align}\label{eq:fidelity in convexity}
    F(\omega^{C N Q R} ,\tau^{\hat{C} \hat{N} \hat{Q} R} )
      &= F(\omega^{C N Q R} ,\lambda \tau_1^{\hat{C} \hat{N} \hat{Q} R}
        + (1-\lambda) \tau_2^{\hat{C} \hat{N} \hat{Q} R}) \nonumber \\
      &= F(\lambda \omega^{C N Q R}+(1-\lambda)\omega^{C N Q R},
           \lambda \tau_1^{\hat{C} \hat{N} \hat{Q} R}
            + (1-\lambda) \tau_2^{\hat{C} \hat{N} \hat{Q} R}) \nonumber\\
      &\geq \lambda F( \omega^{C N Q R},\tau_1^{\hat{C} \hat{N} \hat{Q} R})
            +(1-\lambda)F( \omega^{C N Q R},\tau_2^{\hat{C} \hat{N} \hat{Q} R}) \nonumber\\
     &\geq 1-\left( \lambda\epsilon_1 +(1-\lambda)\epsilon_2 \right).
  \end{align}
  The first inequality is due to simultaneous concavity of the fidelity in both
  arguments;
  the last line follows by the definition of the isometries $U_1$ and $U_2$.
  Thus, the isometry $U_0$ yields a fidelity of at least 
  $1-\left( \lambda\epsilon_1 +(1-\lambda)\epsilon_2 \right) =: 1-\epsilon$.
  Now let $E'=E FF'$ denote the environment of the isometry $U_0$ defined above. 
  According to Definition \ref{def:J_epsilon Z_epsilon}, we obtain
  \begin{align}
    Z_\epsilon(\omega) &\geq S(\hat{N} E'|C')_{\tau} \nonumber\\ 
                     &= S(\hat{N} EFF'|C')_{\tau} \nonumber\\ 
                     &= S(F|C')_{\tau}+S(\hat{N} E|F C')_{\tau}+S(F'|\hat{N} EF C')_{\tau} \label{eq:Z_concavity_1}\\
                     &\geq S(\hat{N}E|FC')_{\tau} \label{eq:Z_concavity_2}\\
                     &= \lambda S(\hat{N} E|C')_{\tau_1}+(1-\lambda) S(\hat{N} E|C')_{\tau_2}\label{eq:Z_concavity_3}\\
                     &= \lambda Z_{\epsilon_1}(\omega)+(1-\lambda)Z_{\epsilon_2}(\omega) \label{eq:Z_concavity_4},
  \end{align}
  where the state $\tau$ in the entropies is given in Eq.~(\ref{eq: tau in convexity proof});
  Eq.~(\ref{eq:Z_concavity_1}) is due to the chain rule; 
  Eq.~(\ref{eq:Z_concavity_2}) follow because 
  %the register $C'$ in state $\omega$ is classical and can be copies to another register $a^{\zegund}_0$, so 
  for the state on systems $\hat{N} EFF' C' $ we have $S(F'|C')+S(F'|\hat{N} E F C')\geq 0$ 
  which follows from strong sub-additivity of the entropy; 
  Eq.~(\ref{eq:Z_concavity_3}) follows by expanding the conditional entropy on the classical system $F$; 
  Eq.~(\ref{eq:Z_concavity_4}) follows from the definitions of the isometries $U_1$ and $U_2$.

  Moreover, let $U_1:C N Q \hookrightarrow \hat{C} \hat{N} \hat{Q} E$ and 
  $U_2:C N Q \hookrightarrow \hat{C} \hat{N} \hat{Q} E$ be the isometries attaining the 
  maximum for $\epsilon_1$ and $\epsilon_2$ in the definition of $J_{\epsilon}(\omega)$, respectively.
  Again, define the isometry $U_0$ as in Eq.~(\ref{eq: isometry U in convexity}),
  which leads to the bound on the fidelity as in Eq.~(\ref{eq:fidelity in convexity}),
  letting $E'=EFF'$ be the environment of the isometry $U_0$. 
  According to Definition \ref{def:J_epsilon Z_epsilon}, we obtain
  \begin{align}
    J_\epsilon(\omega) &\geq I(\hat{N} E FF':\hat{C} \hat{Q}|C')_{\tau} \nonumber \\
                     &\geq I(\hat{N} E F:\hat{C} \hat{Q}|C')_{\tau} \label{eq:concavity_J_1} \\
                     &= I(F:\hat{C} \hat{Q}|C')_{\tau}+I(\hat{N} E :\hat{C} \hat{Q}|F C')_{\tau} \label{eq:concavity_J_2} \\
                     &\geq I(\hat{N} E :\hat{C} \hat{Q}|F C')_{\tau} \label{eq:concavity_J_3} \\
                     &=    \lambda I(\hat{N} E :\hat{C} \hat{Q}| C')_{\tau_1}+(1-\lambda) I(\hat{N} E :\hat{C} \hat{Q}| C')_{\tau_2} \label{eq:concavity_J_4}\\
                     &=    \lambda J_{\epsilon_1}(\omega)+(1-\lambda)J_{\epsilon_2}(\omega)\label{eq:concavity_J_5},
  \end{align}
  where Eq.~(\ref{eq:concavity_J_1}) follows from data processing; 
  Eq.~(\ref{eq:concavity_J_2}) is due to the chain rule for mutual information;
  Eq.~(\ref{eq:concavity_J_3}) follows from strong sub-additivity of the 
  entropy, $I(F:\hat{C} \hat{Q}|C')_{\tau} \geq 0$;
  Eq.~(\ref{eq:concavity_J_4}) is obtained by expanding the conditional mutual 
  information  on the classical system $F$; 
  finally, Eq.~(\ref{eq:concavity_J_5}) follows from the definitions of the isometries $U_1$ and $U_2$.

%%%%%%%%%%%%%%%%%% 3 %%%%%%%%%%%%%%%%%%%%
\item The functions are non-decreasing and concave for $\epsilon \geq 0 $, so they are continuous 
  for $\epsilon > 0$. 
  The concavity implies furthermore that $J_{\epsilon}$ and $Z_{\epsilon}$ are lower semi-continuous at 
  $\epsilon=0$. On the other hand, since the fidelity, the conditional entropy and the conditional 
  mutual information are all continuous functions of CPTP maps, and the domain of both optimizations 
  is a compact set, we conclude that $J_\epsilon(\omega)$ and $Z_{\epsilon}$ are also upper 
  semi-continuous at $\epsilon=0$, so they are continuous at $\epsilon=0$ 
  \cite[Thms.~10.1 and 10.2]{Rockafeller}. 

%%%%%%%%%%%%%%%%%% 4 %%%%%%%%%%%%%%%%%%%%
\item We first prove 
  $Z_{\epsilon}(\omega_1 \otimes \omega_2) \leq  Z_{\epsilon}(\omega_1) +Z_{\epsilon}(\omega_2)$.
  In the definition of $Z_{\epsilon}(\omega_1 \otimes \omega_2)$, let the isometry 
  $U_0:C_1 N_1 Q_1 C_2 N_2 Q_2 \hookrightarrow \hat{C}_1 \hat{N}_1 \hat{Q}_1 \hat{C}_2 \hat{N}_2 \hat{Q}_2 E$
  be the one attaining the maximum, which acts on the following purified source states with purifying 
  systems $R'_1$ and $R'_2$: 
  \begin{equation}
    \label{eq:U0-action}
    \ket{\tau}^{\hat{C}_1 \hat{N}_1 \hat{Q}_1 \hat{C}_2 \hat{N}_2 \hat{Q}_2 E R_1 C'_1 R'_1 R_2 C'_2 R'_2}
               =(U_0 \otimes \1_{R_1 C'_1 R'_1 R_2 C'_2 R'_2})\ket{\omega_1}^{C_1 N_1 Q_1 R_1 C'_1 R'_1}
                                                   \otimes    \ket{\omega_2}^{C_2 N_2 Q_2 R_2 C'_2 R'_2}.
  \end{equation}
  By definition, the fidelity is bounded by
  \begin{align*}
    F(\omega_1^{C_1 N_1 Q_1 R_1} \otimes \omega_2^{C_2 N_2 Q_2 R_2},
      \tau^{\hat{C}_1 \hat{N}_1 \hat{Q}_1 \hat{C}_2 \hat{N}_2 \hat{Q}_2 R_1 R_2}) \geq 1- \epsilon.   
  \end{align*}
  Now, we can define an isometry 
  $U_1:C_1 N_1 Q_1 \hookrightarrow \hat{C}_1 \hat{N}_1 \hat{Q}_1 E_1$ 
  acting only on systems $C_1 N_1 Q_1$, by letting
  $U_1 = (U_0 \otimes \1_{R_2 C_2' R_2'})(\1_{C_1 N_1 Q_1} \otimes \ket{\omega_2}^{C_2 N_2 Q_2 R_2 C_2' R_2})$
  and with the environment $E_1 := \hat{C}_2 \hat{N}_2 \hat{Q}_2 E R_2  C'_2 R'_2$.
  It has the property that 
  $\ket{\tau}^{\hat{C}_1 \hat{N}_1 \hat{Q}_1 R_1 C_1' R_1' E} 
   = (U_1 \otimes \1_{R_1 C_1' R_1'})\ket{\omega_1}^{C_1 N_1 Q_1 R_1 C_1' R_1'}$ 
  has the same reduced state on $\hat{C}_1 \hat{N}_1 \hat{Q}_1 R_1$ as $\tau$ from
  Eq. (\ref{eq:U0-action}).
  This isometry preserves the fidelity for $\omega_1$, which follows from monotonicity 
  of the fidelity under partial trace:
  \begin{align*}
     F(\omega_1^{C_1 N_1 Q_1 R_1},\tau_1^{\hat{C}_1 \hat{N}_1 \hat{Q}_1 R_1}) 
       &= F(\omega_1^{C_1 N_1 Q_1 R_1},\tau^{\hat{C}_1 \hat{N}_1 \hat{Q}_1 R_1}) \\
       &\geq F(\omega_1^{C_1 N_1 Q_1 R_1} \otimes \omega_2^{C_2 N_2 Q_2 R_2},
               \tau^{\hat{C}_1 \hat{N}_1 \hat{Q}_1 \hat{C}_2 \hat{N}_2 \hat{Q}_2 R_1 R_2}) \\
       &\geq 1- \epsilon.   
  \end{align*}
  By the same argument, there is an isometry 
  $U_2:C_2 N_2 Q_2\hookrightarrow \hat{C}_1 \hat{N}_1 \hat{Q}_1 \hat{C}_2 \hat{N}_2 \hat{Q}_2 E R_1 C'_1 R'_1$
  with output system $\hat{C}_2 \hat{N}_2 \hat{Q}_2$ and 
  environment $E_2:=\hat{C}_1 \hat{N}_1 \hat{Q}_1 E R_1 C'_1 R'_1$, such that
  \begin{align*}
    F(\omega_2^{C_2 N_2 Q_2 R_2},\tau_2^{\hat{C}_2 \hat{N}_2 \hat{Q}_2 R_2}) 
      &=    F(\omega_2^{C_2 N_2 Q_2 R_2},\tau^{\hat{C}_2 \hat{N}_2 \hat{Q}_2 R_2}) \\
      &\geq F(\omega_1^{C_1 N_1 Q_1 R_1} \otimes \omega_2^{C_2 N_2 Q_2 R_2},
              \tau^{\hat{C}_1 \hat{N}_1 \hat{Q}_1 \hat{C}_2 \hat{N}_2 \hat{Q}_2 R_1 R_2}) \\
      &\geq 1- \epsilon.   
  \end{align*}
  Therefore, we obtain:
  \begin{align}
     Z_{\epsilon}(\omega_1) &+Z_{\epsilon}(\omega_2)-Z_{\epsilon}(\omega_1 \otimes \omega_2) \nonumber\\
     &\geq
     S(\hat{N}_1 E_1|C'_1)_{\tau}+S(\hat{N}_2 E_2|C'_2)_{\tau}-S(\hat{N}_1
     \hat{N}_2 E |C'_1 C'_2)_{\tau} \label{eq:Z_additivity_1}\\
     &=S(\hat{N}_1 E_1 C'_1)_{\tau}+S(\hat{N}_2 E_2C'_2)_{\tau}-S(\hat{N}_1
     \hat{N}_2 E C'_1 C'_2)_{\tau}-S(C'_1)-S(C'_2)+S(C'_1 C'_2) \label{eq:Z_additivity_2}\\
     &=S(\hat{N}_1 E_1 C'_1)_{\tau}+S(\hat{N}_2 E_2C'_2)_{\tau}-S(\hat{N}_1
     \hat{N}_2 E C'_1 C'_2)_{\tau} \label{eq:Z_additivity_3}\\
     &=S(\hat{C}_1\hat{Q}_1 R_1 R'_1)+S(\hat{C}_2\hat{Q}_2 R_2 R'_2)-S(\hat{C}_1\hat{Q}_1 \hat{C}_2\hat{Q}_2 R_1 R'_1 R_2 R'_2) \label{eq:Z_additivity_4}\\
     &=I(\hat{C}_1\hat{Q}_1 R_1 R'_1:\hat{C}_2\hat{Q}_2 R_2 R'_2) \nonumber\\
     &\geq 0 \label{eq:Z_additivity_5},
  \end{align}
  where Eq.~(\ref{eq:Z_additivity_1}) is due to Definition~\ref{def:J_epsilon Z_epsilon};
  Eq.~(\ref{eq:Z_additivity_2}) is due to the chain rule;
  Eq.~(\ref{eq:Z_additivity_3}) because the systems $C'_1$ and $C'_2$ are independent from each other;
  Eq.~(\ref{eq:Z_additivity_4}) follows because the overall state on systems 
  $\hat{C}_1 \hat{N}_1 \hat{Q}_1 \hat{C}_2 \hat{N}_2 \hat{Q}_2 E R_1 C'_1 R'_1 R_2 C'_2 R'_2$ 
  is pure;
  Eq.~(\ref{eq:Z_additivity_5}) is due to sub-additivity of the entropy. 

  To prove prove 
  $J_{\epsilon}(\omega_1 \otimes \omega_2) \leq J_{\epsilon}(\omega_1) +J_{\epsilon}(\omega_2)$,
  let the isometry 
  $U_0:C_1 N_1 Q_1 C_2 N_2 Q_2 \hookrightarrow \hat{C}_1 \hat{N}_1 \hat{Q}_1 \hat{C}_2 \hat{N}_2 \hat{Q}_2 E$ 
  be the one attaining the maximum in definition of $J_{\epsilon}(\omega_1 \otimes \omega_2)$,
  which acts on the following purified source states with purifying 
  systems $R'_1$ and $R'_2$, as in Eq. (\ref{eq:U0-action}).
  By definition, the fidelity is bounded as
  \begin{align*}
    F(\omega_1^{C_1 N_1 Q_1 R_1} \otimes \omega_2^{C_2 N_2 Q_2 R_2},
      \tau^{\hat{C}_1 \hat{N}_1 \hat{Q}_1 \hat{C}_2 \hat{N}_2 \hat{Q}_2 R_1 R_2}) 
               \geq 1- \epsilon.   
  \end{align*}
  Now define 
  $U_1:C_1 N_1 Q_1\hookrightarrow \hat{C}_1 \hat{N}_1 \hat{Q}_1 \hat{C}_2 \hat{N}_2 \hat{Q}_2 E R_2  C'_2 R'_2$ 
  and $U_2:C_2 N_2 Q_2\hookrightarrow \hat{C}_1 \hat{N}_1 \hat{Q}_1 \hat{C}_2 \hat{N}_2 \hat{Q}_2 E R_1 C'_1 R'_1$ 
  as in the above discussion, with the environments 
  $E_1:=\hat{C}_2 \hat{N}_2 \hat{Q}_2 E R_2 C'_2 R'_2$ and 
  $E_2:=\hat{C}_1 \hat{N}_1 \hat{Q}_1 E R_1 C'_1 R'_1$, respectively. 
  Recall that the fidelity for the states $\omega_1$ and $\omega_2$ is at least 
  $1-\epsilon$, because of the monotonicity of the fidelity under partial trace. 
  Thus we obtain
  \begin{align}
  J_{\epsilon}(\omega_1) 
     &+J_{\epsilon}(\omega_2)-J_{\epsilon}(\omega_1 \otimes \omega_2) \nonumber\\
     &\geq I(\hat{N}_1 E_1:\hat{C}_1\hat{Q}_1|C'_1)_\tau+
       I(\hat{N}_2 E_2:\hat{C}_2\hat{Q}_2|C'_2)_\tau-I(\hat{N}_1\hat{N}_2 E:\hat{C}_1\hat{Q}_1\hat{C}_2\hat{Q}_2|C'_1 C'_2)_\tau
       \label{eq:J_additivity_1}\\
     &=S(\hat{N}_1 E_1 C'_1)+S(\hat{C}_1\hat{Q}_1C'_1)-S( \hat{C}_1 \hat{N}_1 \hat{Q}_1 E_1 C'_1)-S(C'_1) \nonumber \\
     &\quad +S(\hat{N}_2 E_2 C'_2)+S(\hat{C}_2\hat{Q}_2C'_2)-S(     \hat{C}_2 \hat{N}_2 \hat{Q}_2 E_2 C'_2)-S(C'_2) \nonumber\\
     &\quad \!-\!S(\hat{N}_1\hat{N}_2 E C'_1 C'_2) \!- \!S(\hat{C}_1\hat{Q}_1\hat{C}_2\hat{Q}_2 C'_1 C'_2)\!+\!S( \hat{C}_1\hat{N}_1\hat{Q}_1\hat{C}_2 \hat{N}_2\hat{Q}_2E C'_1 C'_2)\!+\! S(C'_1 C'_2) \label{eq:J_additivity_2} \\
     &=S(\hat{C}_1 \hat{Q}_1 R_1 R'_1)+S(\hat{C}_1\hat{Q}_1C'_1)-S(R_1 R'_1)-S(C'_1) \nonumber \\
     &\quad +S(\hat{C}_2 \hat{Q}_2 R_2 R'_2)+S(\hat{C}_2\hat{Q}_2C'_2)-S(     R_2 R'_2)-S(C'_2) \nonumber\\
     &\quad \!-\!S(\hat{C}_1 \hat{Q}_1\hat{C}_2 \hat{Q}_2 R_1 R'_1 R_2 R'_2) \!- \!S(\hat{C}_1\hat{Q}_1\hat{C}_2\hat{Q}_2 C'_1 C'_2)\!+\!S(R_1 R'_1 R_2 R'_2)\!+\! S(C'_1 C'_2) \label{eq:J_additivity_3}\\
     &=I(\hat{C}_1 \hat{Q}_1 R_1 R'_1:\hat{C}_2 \hat{Q}_2 R_2 R'_2)
     -I(R_1 R'_1:R_2 R'_2)
     +I(\hat{C}_1\hat{Q}_1C'_1:\hat{C}_2\hat{Q}_2C'_2)
     -I(C'_1:C'_2) \nonumber \\
     &\geq I( R_1 R'_1:R_2 R'_2)
     -I(R_1 R'_1:R_2 R'_2)
     +I(C'_1:C'_2)
     -I(C'_1:C'_2) \label{eq:J_additivity_4}\\
     &=0, \nonumber
  \end{align}
  where Eq.~(\ref{eq:J_additivity_1}) is due to Definition~\ref{def:J_epsilon Z_epsilon}; 
  In Eq.~(\ref{eq:J_additivity_2}) we expand the mutual informations in terms of entropies;
  Eq.~(\ref{eq:J_additivity_3}) follows because the overall state on systems
  $\hat{C}_1 \hat{N}_1 \hat{Q}_1 \hat{C}_2 \hat{N}_2 \hat{Q}_2 E R_1 C'_1 R'_1 R_2 C'_2 R'_2$
  is pure; 
  Eq.~(\ref{eq:J_additivity_4}) is due to data processing. 

%%%%%%%%%%%%%%%%%% 5 %%%%%%%%%%%%%%%%%%%%
\item According to Theorem~\ref{thm: KI decomposition} \cite{KI2002,Hayden2004}, 
  any isometry $U:C N Q \rightarrow \hat{C} \hat{N} \hat{Q} E$ acting on the state 
  $\omega^{C N Q R C'}$ which preserves the reduced state on systems $C N Q R C'$ 
  ($C'$ here is considered as a part of the reference system), acts as the following:
  \begin{align*}
    (U \otimes \1_{RC'}) \omega^{C N Q R C'}(U^{\dagger} \otimes \1_{RC'})
      =\sum_{j} p_j \proj{j}^{C}\otimes U_j \omega_j^{N} U_j^{\dagger} \otimes \rho_{j}^{Q R} \otimes \proj{j}^{C'},
  \end{align*}
  where the isometry $U_j: N \rightarrow \hat{N} E$ satisfies 
  $\Tr_E [U_j \omega_j^{N} U_j^{\dagger}]=\omega_j$.
  Therefore,  in Definition~\ref{def:J_epsilon Z_epsilon} for $\epsilon=0$, the final state is
  \begin{align*}
    \tau^{\hat{C} \hat{N} \hat{Q}  E R C'}
      = \sum_{j} p_j \proj{j}^{C}\otimes U_j \omega_j^{N} U_j^{\dagger} \otimes \rho_{j}^{Q R} \otimes \proj{j}^{C'}.
  \end{align*}
  Thus we can directly evaluate
\begin{align*}
      Z_0(\omega)=S(\hat{N} E|C')_\tau=S(N |C)_\omega \text{ and } 
      J_0(\omega)=I(\hat{N} E:\hat{C}\hat{Q}|C')_\tau=0,
\end{align*}
concluding the proof.
\hfill\qedsymbol
\end{enumerate}

\section{Discussion}
\label{sec:Discussion}
We have introduced a common framework for all single-source quantum compression 
problems, i.e. settings without side information at the encoder or the decoder, 
by defining the compression task as the reproduction of a given bipartite state 
between the system to be compressed and a reference. That state, which defines 
the task, can be completely general, and special instances recover Schumacher's
quantum source compression (in both variants of a pure state ensemble and of 
a pure entangled state) \cite{Schumacher1995} 
and compression of a mixed state ensemble source in the blind variant 
\cite{Horodecki1998,KI2001}.

Our general result gives the optimal quantum compression rate in terms of
qubits per source, both in the settings without and with entanglement, and 
indeed the entire qubit-ebit rate region, reproducing the aforementioned 
special cases, along with other previously considered problems \cite{ZBK2019}. 
Despite the technical difficulties in obtaining it, the end result has a 
simple and intuitive interpretation. Namely, the given source $\rho^{AR}$ 
is equivalent to a source in standard Koashi-Imoto form,
\[
  \omega^{CQR} = \sum_j p_j \proj{j}^C \otimes \rho_j^{QR},
\]
so that $j$ has to be compressed as classical information, at rate $S(C)$,
and $Q$ as quantum information, at rate $S(Q|C)$; in the presence of 
entanglement, the former rate is halved while the latter is maintained. 
Indeed, what our Theorem \ref{theorem:complete rate region}
shows is that the original source has the same qubit-ebit 
rate region as the clean classical-quantum mixed source
\[
  \Omega^{CQRR'C'} = \sum_j p_j \proj{j}^C \otimes \proj{\psi_j}^{QRR'} \otimes \proj{j}^{C'},
\]
where $\ket{\psi_j}^{QRR'}$ purifies $\rho_j^{QR}$, and $RR'C'$ is considered
the reference. In $\Omega$, $C$ is indeed a manifestly classical source, 
since it is duplicated in the reference system, and conditional on $C$,
$Q$ is a  genuinely quantum source since it is purely entangled with the
reference system. As $\Tr_{R'C'} \Omega^{CQRR'C'} = \omega^{CQR}$, any 
code and any achievable rates for $\Omega$ are good for $\omega$, and 
that is how the achievability of the rate region in Theorem \ref{theorem:complete rate region} 
can be described. The opposite, that a code good for $\omega$ should be 
good for $\Omega$, is far from obvious. Indeed, if that were true, it would 
not only yield a quick and simple proof of our converse bounds, but would 
imply that the rate region of Theorem \ref{theorem:complete rate region} satisfies a 
strong converse! However, as we do not know this reduction to the source $\Omega$,
our converse proceeds via a more complicated, indirect route, and yields only
a weak converse. Whether the strong converse holds, and what the detailed
relation between the sources $\omega^{CQR}$ and $\Omega^{CQRR'C'}$ is, 
remain open questions. 

\medskip
As we were finishing the write-up of the present paper, we became aware of 
related work by Anshu \emph{et al.} \cite{AnshuLeungTouchette}, who consider 
a source consisting of a commuting mixed state ensemble, with the aim of 
showing a large separation between the Holevo information of the ensemble 
and the actual (blind) compression rate of the ensemble, even at non-zero
error. Unlike our work, which follows source coding convention by considering
block error, they define the error as ``error (infidelity) per letter'', which
is a weaker requirement, and prove a rate lower bound in their 
\cite[Thm. 2]{AnshuLeungTouchette}.
It is worth noting that first, our lower bounds in Theorem \ref{theorem:complete rate region}
only require the error per letter criterion, and that indeed Eqs. (\ref{eq:converse_Q})
and (\ref{eq:converse_Q+E}) give rate lower bounds for asymptotically large $n$ 
and non-zero error $\epsilon$, which in addition in the limit $\epsilon \rightarrow 0$
become tight: 
\begin{align*}
  Q   &\geq S(C Q)-\frac12 S(C) 
                  -\frac12 J_{\epsilon}(\omega)
                  -\frac12 \bigl(Z_{\epsilon}(\omega)-S(N|C)\bigr) - \frac12 \delta(\epsilon), \\
  Q+E &\geq S(C Q) - J_{\epsilon}(\omega) - \delta(\epsilon),
\end{align*}
where $J_\epsilon$ and $Z_\epsilon$ are as in Definition \ref{def:J_epsilon Z_epsilon},
and $\delta(\epsilon) = \sqrt{2\epsilon} \log(|C||Q|)$.

%\section{For myself ZBK: Notation and other}
%$\cE: A^n C \longrightarrow C $\\
% 
%$\cD:C B_0 \longrightarrow \hat{A}^n$\\ 
%
%$\cR:C N Q \longrightarrow A$\\ 
%
%$U_{\cE} : A^n C \hookrightarrow C W$\\
%
%$U_{\cD} : C B_0 \hookrightarrow \hat{A}^n V$\\
%
%$U_{\cR} : C N Q \hookrightarrow A Y$\\
%
%$U_{\KI} : A \hookrightarrow C N Q$\\
%
%$U_{\KI}^{\otimes n} : \hat{A}^n \hookrightarrow \hat{C}^n \hat{N}^n \hat{Q}^n$\\
%
%VWY .  EF .   U for isometries
%
%$\omega^{C N Q C' R}=\sum_{j} p_j \proj{j}^{C}\otimes \omega_j^{N} \otimes \rho_{j}^{Q R} \otimes \proj{j}^{C'}$
%
%Show all isometries with U.\\

%%%%%%%%%%%%%%%%%%%%%%%%%%%%%%%%%%%%%%%%%%%%%%%%%%%%%%%%%%%%%%%%%%%%%%%%%%%%%%%%%%%%%%%%%%%%%%%

\bigskip\bigskip
\textbf{Acknowledgments.}
The authors were supported by the Spanish MINECO (projects 
FIS2016-80681-P, FISICATEAMO FIS2016-79508-P and SEVERO OCHOA No. SEV-2015-0522, FPI) 
with the support of FEDER funds, the Generalitat de Catalunya (projects 2017-SGR-1127, 
2017-SGR-1341 and CERCA/Program), ERC AdG OSYRIS, EU FETPRO QUIC, 
and the National Science Centre, Poland-Symfonia grant no. 2016/20/W/ST4/00314.

%%%%%%
%% To balance the columns at the last page of the paper use this
%% command:
%%
%\enlargethispage{-1.2cm} 
%%
%% If the balancing should occur in the middle of the references, use
%% the following trigger:
%%
%\IEEEtriggeratref{3}
%%
%% which triggers a \newpage (i.e., new column) just before the given
%% reference number. Note that you need to adapt this if you modify
%% the paper.  The "triggered" command can be changed if desired:
%%
%\IEEEtriggercmd{\enlargethispage{-20cm}}
%%
%%%%%%

%%%%%%
%% References:
%% We recommend the usage of BibTeX:
%%
%\bibliographystyle{IEEEtran}
%\bibliography{definitions,bibliofile}
%%
%% where we here have assume the existence of the files
%% definitions.bib and bibliofile.bib.
%% BibTeX documentation can be obtained at:
%% http://www.ctan.org/tex-archive/biblio/bibtex/contrib/doc/
%%%%%%

%% Or you use manual references (pay attention to consistency and the formatting style!):

%\vfill

\bibliographystyle{IEEEtran}

\end{document}